\documentclass[aps,prb,showpacs,twocolumn,floatfix]{revtex4}

\usepackage{graphicx}
\usepackage{bm}
\usepackage{amssymb}
\usepackage{amsmath}
\input epsf

\begin{document}

\title{Stability of conductance oscillations in monatomic sodium wires}
\author{Petr A. Khomyakov}\thanks{Corresponding author}\email[E-mail addess: ]{p.khomyakov@utwente.nl}
\homepage{http://cms.tnw.utwente.nl} \author{Geert Brocks}
 \affiliation{Computational
Materials Science, Faculty of Science and Technology and MESA+
 Institute for Nanotechnology, University of Twente, P.O. Box 217, 7500 AE Enschede,
The Netherlands.}
\date{\today}

\begin{abstract}
We study the stability of conductance oscillations in monatomic
sodium wires with respect to structural variations. The geometry,
the electronic structure and the electronic potential of sodium
wires suspended between two sodium electrodes are obtained from
self-consistent density functional theory calculations. The
conductance is calculated within the framework of the
Landauer-B\"utttiker formalism, using the mode-matching technique as
formulated recently in a real-space finite-difference scheme [Phys.
Rev. B \textbf{70}, 195402 (2004)]. We find a regular even-odd
conductance oscillation as a function of the wire length, where
wires comprising an odd number of atoms have a conductance close to
the quantum unit $G_0=e^2/\pi\hbar$, and even-numbered wires have a
lower conductance. The conductance of odd-numbered wires is stable
with respect to geometry changes in the wire or in the contacts
between the wire and the electrodes; the conductance of
even-numbered wires is more sensitive. Geometry changes affect the
spacing and widths of the wire resonances. In the case of
odd-numbered wires the transmission is on-resonance, and hardly
affected by the resonance shapes, whereas for even-numbered wires
the transmission is off-resonance and sensitive to the resonance
shapes. Predicting the amplitude of the conductance oscillation
requires a first-principles calculation based upon a realistic
structure of the wire and the leads. A simple tight-binding model is
introduced to clarify these results.
\end{abstract}  \maketitle

\section{Introduction}
Recent progress in fabricating conductors of atomic dimensions has
stimulated a large number of experimental and theoretical studies on
these nanoscale devices.\cite{Agrait:prp03} Conductors whose cross
section contains only a small number of atoms are commonly called
``atomic wires''. Clear evidence that the fundamental limit of a one
atom cross section can be reached, has been presented for gold
atomic wires.\cite{Yanson:nat98,Ohnishi:nat98} Over the last decade
the electronic transport in atomic wires made of various metals has
been characterized in great detail experimentally.
\cite{Krans:nat95,Yazdani:sc96,Yanson:nat99,
Rodrigues:prl00,Smit:prl03,Csonka:prl03} Such wires have
conductances of the order of the quantum unit $G_{0}=e^2/\pi\hbar$,
so the description of their transport properties, as well as of
their atomic and electronic structures, requires a full
quantum-mechanical treatment.\cite{Datta:95}

Simple theoretical schemes have been proposed, in which the atomic
wire is described by a jellium\cite{Stafford:prl97} or a
tight-binding model.\cite{Emberly:prb98} At present a
first-principles approach based on density functional theory (DFT)
gives the most advanced description of the geometry and electronic
structure of atomic wires. Several theoretical methods that combine
the Landauer-B\"uttiker formalism with DFT, have been developed in
order to solve the quantum transport problem in terms of scattering
amplitudes.\cite{Hirose:prl95,Lang:prb95,Choi:prb99,Fujimoto:prb03,
Khomyakov:prb04,Xia:prb06} Alternatively, a Green's function
formalism is commonly used for solving the transport
problem.\cite{Xue:jchp01,Damle:prb01,Taylor:prb01,Nardelli:prb01,Palacios:prb02,Dreher:prb05,
Brandbyge:prb02,Wortmann:prb02b,Thygesen:prb03,Bagrets:prb04,
Havu:prb06,Li:jp06,Major:cond05} Both these approaches are in fact
completely equivalent in the case of noninteracting
electrons.\cite{Khomyakov:prb05}

Atomic wires that have a cross section of just one atom, so-called
``monatomic'' wires, are the ultimate examples of
quasi-one-dimensional systems. Here the effects of a reduced
dimensionality are expected to be most pronounced. {\it A priori}
the existence of monatomic wires is not obvious. Such free-standing
one-dimensional structures might be unstable because of the low
coordination number of the atoms in the wire. Molecular dynamics
simulations based upon an effective medium
model\cite{Sorensen:prb98} or a tight-binding
model\cite{Silva:prl01,Dreher:prb05} have been used to study the
stability of a wire as a function of its elongation. Since such
simulations use highly simplified interatomic potentials, they aim
at providing a qualitative understanding of the wire formation. A
more quantitative description can be provided by first-principles
DFT
calculations,\cite{Barnett:nat97,Sanchez:prl99,Okamoto:prb99,Hakkinen:prb00,Rubio:prl01}
but then only relatively small systems can be handled. A multitude
of different structures has been studied, such as dimerized, zigzag,
and helical
wires.\cite{DeMaria:chpl00,Sanchez:sursc01,Sen:prb01,Springborg:prb03,
Ono:prb03,Asaduzzaman:prb05}

One of the most striking features of monatomic wires is the
nonmonotonic behavior of the conductance as a function of the number
of atoms in the wire.\cite{Yamaguchi:ssc97,Lang:prl97} Such a
behavior has been predicted by Lang for wires consisting of
monovalent atoms.\cite{Lang:prl97} His model assumes a chain of
atoms suspended between two planar semi-infinite jellium electrodes.
The conductance predicted by this model is much lower than the
quantum unit, which disagrees with experiments on monovalent atomic
chains.\cite{Krans:nat95,Yanson:nat98,Yanson:nat99} However, the
model can be modified in a simple way by adding a basis consisting
of three atoms on top of the jellium
electrodes.\cite{Kobayashi:prb00} This reduces the charge transfer
between the wire and leads and it reduces spurious reflections at
the wire-electrode interfaces. The conductance of a one-atom wire is
then close to the quantum unit, in agreement with experiment.

From the Friedel sum rule it can be shown that the conductance of an
atomic chain consisting of monovalent atoms exhibits a regular
oscillation with respect to the number of atoms in the
chain.\cite{Sim:prl01} Moreover, assuming mirror reflection and
time-reversal symmetries together with local charge neutrality of
the wire, the conductance of wires with an odd number of atoms is
expected to be very close to the quantum unit.\cite{Lee:prb01} The
period of the conductance oscillation then equals two atoms. This
has been confirmed in conductance calculations for wires connected
to jellium leads via an atomic basis.\cite{Tsukamoto:prb02}

The oscillating behavior of the conductance as a function of the wire length has been observed
experimentally in gold wires,\cite{Smit:prl03} and a small parity effect has recently been found in
silver wires.\cite{Thijssen:prl06} Oscillations have also been observed in wires consisting of
atoms with a higher valency.\cite{Yanson:nat98,Thygesen:prl03}

Since the discovery of the parity effect in wires of monovalent
atoms there has been a discussion on how sensitive the conductance
oscillation is to the geometry of the wire and the contacts. The
general arguments given in Ref.~\onlinecite{Sim:prl01} suggest that
the conductance for odd-numbered wires should always be higher than
for even-numbered ones, provided that the wires are locally charge
neutral. For a sodium wire connected to (artificial) \textit{fcc}
sodium electrodes the charge transfer has been estimated and it is
found to be rather small.\cite{LeeYJ:prb04} Conductance calculations
based upon a scattering approach have recently been performed for
short sodium wires attached to sodium electrodes with a more
realistic \textit{bcc} structure.\cite{Khomyakov:prb04,Egami:prb05}
The phase of the conductance oscillation obtained in these
calculations, is consistent with that found in the jellium electrode
calculations.\cite{Sim:prl01,Tsukamoto:prb02}

Other studies predict however that the conductance oscillation found
in wires of monovalent atoms is very sensitive to the geometry. Even
the phase of the oscillation can be reversed such, that
even-numbered wires have a larger conductance than odd-numbered
ones.\cite{Gutierrez:appb01,Havu:prb02,Viljas:prb05,Major:cond05}.
Using the Friedel sum rule to calculate the conductance of a wire
connected to jellium electrodes it has been found that the phase of
the conductance oscillation is reversed if the jellium leads become
sufficiently sharp.\cite{Havu:prb02} This has not been confirmed by
later calculations using a scattering approach to calculate the
conductance, which give results that are consistent with Lang's
findings for planar jellium electrodes.\cite{Havu:prb06,Lang:prl97}

From strictly one-dimensional linear combination of atomic orbitals
DFT (LCAO-DFT) calculations, i.e. sodium chains coupled to
one-dimensional metallic leads, it has been argued that there is a
critical distance between the wire and the leads where the
conductance oscillation changes its phase and even-numbered chains
become more conductive than odd-numbered
ones.\cite{Gutierrez:appb01} A change of phase has also been
predicted to occur upon elongating the wire by adding atoms. In
short wires the odd-numbered chains then have the higher conductance
and in long wires the even-numbered ones have the higher
conductance. Similar effects have also been claimed recently for
particular atomic configurations in calculations using
three-dimensional leads.\cite{Major:cond05,Viljas:prb05}

In conclusion, some of the results that appeared in the literature regarding the even-odd
conductance oscillation in monatomic wires seem to be contradictory. In this paper we present the
results of conductance calculations for monatomic sodium wires in order to investigate the effect
of the wire geometry and the wire-lead coupling on the conductance oscillation. Since sodium has a
simple electronic structure, a sodium wire is one of the simplest examples of an atomic wire. As
such it is an important reference system for studying wires with a more complicated electronic
structure, and it can be used as a system for benchmarking theoretical and computational
techniques. We perform first-principles conductance calculations based on the mode-matching
technique\cite{Khomyakov:prb04} on sodium wires suspended between sodium electrodes, while
systematically varying the atomic configuration of the wire and that of the wire-lead contacts. The
entire system consisting of the wire and the semi-infinite electrodes, is treated fully
atomistically.

We find that the parity effect, i.e. the even-odd conductance
oscillation, is very robust with respect to changing the structure
of the wire and to varying the strength of the coupling between the
wire and the leads. In the conductance of long wires we find no
tendency to a phase change in the even-odd oscillation. The
conductance is analyzed using the electronic levels of free-standing
wires in order to interpret the parity effect in terms of
transmission resonances. In addition, we analyze our
first-principles results using a simple tight-binding model. In
particular, we show that local charge neutrality of the sodium wires
provides a strong constraint on the phase of the conductance
oscillation for all atomic structures considered. In absence of a
significant charge transfer between the wire and the leads, a
transmission resonance is pinned at the Fermi energy for wires
containing an odd number of atoms, which leads to a conductance
close to one quantum unit. Obtaining quantitative values for the
conductance, particularly for even-numbered wires, requires
well-converged first-principles calculations using a realistic
structure of the wire and the leads.

The structure of this paper is as follows. In Sec.~\ref{geometry} we discuss the geometry of
infinite and finite sodium wires. The even-odd oscillation of the conductance is discussed in
general terms in Sec.~\ref{natres}. We investigate the effects on the conductance of varying the
wire geometry and the contacts between wire and leads in Sec.~\ref{univbehav}. Current-voltage
characteristics are analyzed in Sec.~\ref{currvolt}. In Sec.~\ref{otherworks} we compare our
results to those obtained in previous studies. A summary and conclusions are presented in the last
section. The important technical detail of $\mathbf{k}$-point sampling is discussed in
Appendix~\ref{kpoint}.

\section{Structure of sodium wires}\label{geometry}
In this section we investigate possible structures of sodium
monatomic wires by DFT total energy calculations in combination with
geometry optimizations. DFT total energies are calculated with the
PW91 generalized gradient approximation (GGA)
functional\cite{Perdew:prb92} and the projector augmented wave (PAW)
method,\cite{Kresse:prb99,Blochl:prb94} as implemented in the Vienna
\em Ab initio \em Simulation Package
(VASP).\cite{Kresse:prb96,Kresse:cms96,Kresse:prb93} We use a
standard frozen core PAW potential and a plane wave basis set with a
kinetic energy cutoff of $24$~Ry. A Methfessel-Paxton smearing is
applied in integrations over the Brillioun zone with a smearing
parameter $\sigma=0.1$ eV. First we discuss the structure of the
infinite wire, and then that of finite wires connected to
\textit{bcc} sodium electrodes.

\subsection{Infinite wires}
An orthorhombic supercell is used with cell parameters perpendicular to the wire direction equal to
$17$ {\AA}. Parallel to the wire the cell parameter is optimized using 24 $\mathbf{k}$-points to
sample the Brillouin zone along the wire direction for a cell containing two atoms.

First we consider a linear wire geometry; the optimized Na-Na bond length is given in
Table~\ref{tab1}. We have checked that a symmetry breaking in the form of a Peierls distortion, is
negligible for Na-Na bond lengths near the equilibrium interatomic distance, which agrees with
calculations on other monatomic wires.\cite{Sanchez:sursc01,Sen:prb01} Only for sodium wires that
are stretched to interatomic distances larger than $6.91a_0$ a Peierls distortion takes place,
accompanied by a metal-insulator transition. In order to test the accuracy of the calculations we
have also calculated the optimized Na-Na bond length in bulk sodium and in the sodium dimer. The
accuracy is found to be better than 1\% as compared to the experimental values, see
Table~\ref{tab1}.
\begin{table}[!]
\caption{Na-Na nearest neighbor bond length (in $a_0$) for sodium
wires, bulk sodium and the sodium dimer, compared to all-electron
calculations and to experiment. For the zigzag chain also the bond
angle is given.\label{tab1}}
\begin{tabular}{ccccc}
\toprule
  & this work &  all electron   & experiment \\
  \colrule
  linear  &  6.30 &    &     \\
  zigzag  &  6.85 $(57^{\circ})$ &    &     \\
  bulk  &   6.88 & 6.90 (Ref.~\onlinecite{Perdew:prb93})  &  6.91 (Ref.~\onlinecite{Wychoff:64}) \\
 dimer  &  5.88  &  5.85 (Ref.~\onlinecite{Calaminici:jchp99})  & 5.82 (Ref.~\onlinecite{Huber:79})\\
\botrule
\end{tabular}
\end{table}

One-dimensional chains are often unstable with respect to a
deformation in the transverse direction, which results in a zigzag
structure. A linear conformation is preferred if the interatomic
distance exceeds a critical value.\cite{Sanchez:sursc01} Indeed we
find that a sodium wire with a zigzag structure has a lower energy;
its geometry is presented in Table~\ref{tab1}. It is in reasonable
agreement with the one obtained in a previous DFT-LDA
calculation\cite{Bergara:ijqch03}. A bond angle of $\sim 60^{\circ}$
is typically found also in other monatomic
wires.\cite{Asaduzzaman:prb05} Upon stretching the wire a
transformation from a zigzag to a linear geometry takes place as
soon as the interatomic distance in the linear wire becomes
$\gtrsim6.5a_0$. In Ref. \onlinecite{Sanchez:sursc01} a zigzag
structure has been found in gold and copper wires, whereas in
potassium and calcium wires it exists only under compression. The
stability of the zigzag geometry has, therefore, been related to the
presence of directional $d$-bonds in gold and copper. However, our
results show that a sodium wire behaves similarly, which suggests
that the occurrence of a zigzag geometry is not a result of
$d$-bonds only.

Whereas the lowest energy structure is paramagnetic, in
Ref.~\onlinecite{Bergara:ijqch03} two additional local minima have
been found corresponding to ferromagnetic structures with magnetic
moments $\sim$ $0.02\,\mu_{\rm B}$ and $0.12\,\mu_{\rm B}$,
respectively. Our lowest energy (zigzag and linear) structures are
always paramagnetic. A magnetic ordering occurs for zigzag
structures when the Fermi level crosses two energy bands instead of
just one band, but the magnetism disappears rapidly as soon as the
wire is stretched sufficiently. Small magnetic moments have also
been found in calculations on gold wires.\cite{Bergara:ijqch03} No
trace of magnetic effects has been observed in recent conductance
measurements performed in magnetic fields.\cite{Untiedt:prb04} Since
both experiment and theory favor nonmagnetic structures, we will
only consider nonmagnetic sodium wires in the following.

\subsection{Finite wires}
\begin{figure}[!tbp]
\includegraphics[width=6.5cm,keepaspectratio=true]{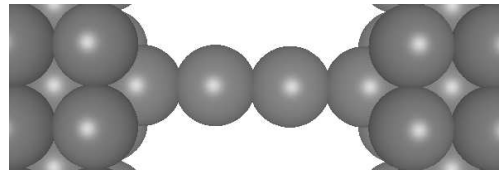}
\caption{(Color online) Supercell representing a two atom sodium
wire between two sodium leads terminated by (001) surfaces. The wire
is connected to electrodes via an apex atom placed on top of each
electrode in a hollow site.} \label{nawire_gopt}
\end{figure}

In this section we discuss the structure of a finite monatomic
sodium wire suspended between two electrodes. A reasonable approach
is to study the structure of the wire near its equilibrium geometry,
which corresponds to the most stable chemical bonding. We use the
equilibrium geometry of the infinite linear wire as a starting point
for finite wires. The electrodes consist of bulk Na in the (001)
orientation. To calculate the structure we use a periodic supercell
that consists of a slab of five layers of sodium for the electrodes.
On top of each electrode surface an apex atom is placed in a hollow
site and a linear wire bridges the two apex atoms as shown in
Fig.~\ref{nawire_gopt}. We use a $12\times 12\times 4$
$\mathbf{k}$-point sampling of the supercell. During the geometry
optimization the atoms in the wire, the apex atoms, and the atoms in
the top surface layer are allowed to relax.

\begin{table}[!]
\caption{Optimized bond lengths (in $a_0$) for a Na wire suspended
between Na electrodes. The columns indicate the number of atoms in a
wire. The row labels $i$-$j$ indicate the distance between the
$i$'th and $j$'th atom in a wire; $\overline{d}$ is the average bond
length. A-1 indicates the distance between the apex atom and the
wire, L1-A the distance between the apex atom and the surface layer,
and L1-L2 the distance between the two top layers of the electrode.
The in-plane Na-Na distance in the top layer is given in the bottom
row.}\label{tab2}
\begin{tabular}{ccccccccccccc}
\toprule
  bonds         & 1    & 2    & 3    & 4    & 5    & 6    & 7    & 8    & 9    \\
\colrule
 1-2            &      & 6.35 & 6.45 & 6.34 & 6.41 & 6.34 & 6.41 & 6.33 & 6.37 \\
 2-3            &      &      & 6.45 & 6.46 & 6.45 & 6.45 & 6.43 & 6.44 & 6.44 \\
 3-4            &      &      &      & 6.34 & 6.45 & 6.36 & 6.41 & 6.37 & 6.40 \\
 4-5            &      &      &      &      & 6.41 & 6.45 & 6.41 & 6.45 & 6.42 \\
 5-6            &      &      &      &      &      & 6.34 & 6.43 & 6.37 & 6.42 \\
 6-7            &      &      &      &      &      &      & 6.41 & 6.44 & 6.40 \\
 7-8            &      &      &      &      &      &      &      & 6.33 & 6.44 \\
 8-9            &      &      &      &      &      &      &      &      & 6.37 \\
 $\overline{d}$ &     & 6.35 & 6.45 & 6.38 & 6.43 & 6.39 & 6.42 & 6.39 & 6.41 \\
 A-1            & 6.53 & 6.59 & 6.51 & 6.54 & 6.49 & 6.51 & 6.47 & 6.48 & 6.46 \\
 L1-A           & 3.72 & 3.64 & 3.60 & 3.58 & 3.52 & 3.52 & 3.45 & 3.43 & 3.38 \\
 L2-L1          & 4.03 & 4.03 & 4.02 & 4.03 & 4.02 & 4.03 & 4.02 & 4.06 & 4.02 \\
 in-plane       & 8.17 & 8.19 & 8.20 & 8.23 & 8.23 & 8.25 & 8.26 & 8.26 & 8.31 \\
 \botrule
 \end{tabular}
\end{table}

The results of the geometry optimization for wires of different lengths are given in
Table~\ref{tab2}. We will discuss the most prominent features of the wire geometries starting from
the electrodes. All structures have mirror symmetry with respect to a plane through the center of
the wire, parallel to the electrode surface. We emphasize that this symmetry is not forced upon the
system, but is the result of the geometry optimization. The top layer of the electrode relaxes
slightly outwards; the distance between the top two layers, L1-L2$\sim 4.03a_0$, is somewhat larger
than the bulk value $3.99a_0$. The distance between the apex atom and the surface L1-A decreases
with the length of the wire, which indicates a growing bond strength. The distance between the apex
atom and the first atom of the wire A-1 is always larger than the maximum bond length between atoms
in the wire. This indicates that bonding within the wire is stronger than bonding to the
electrodes. For gold wires the opposite has been found, i.e. the A-1 distance is shorter than the
average bond length.\cite{Okamoto:prb99}

Focusing upon the interatomic distances between atoms in the wire,
Table~\ref{tab2} clearly shows that even-numbered wires exhibit
dimerization, i.e. an alternation between short and long bonds. A
similar tendency is found in odd-numbered chains, but they have a
topological defect, i.e. a ``kink'', in the center of the wire. The
average bond length of $\sim6.40 a_{0}$ in even and in odd-numbered
wires is larger than the optimized bond length of $6.30a_0$ in the
infinite wire. The infinite chain does not show a dimerization until
the average bond length is larger than $\sim 6.91a_0$ (see the
previous section). This strongly suggests that dimerization in
finite chains is enforced by their boundaries. A qualitatively
similar behavior has also been found in finite gold
wires.\cite{Okamoto:prb99}

Since dimerization of a finite wire is associated with its bonding
to the electrodes, one needs to check how sensitive the dimerization
pattern is to the connection between wire and electrode. We have
performed calculations on larger lateral supercells, and have also
made the connection more graduate by putting a base of four atoms
between the electrode surface and the apex atom. These structural
variations give essentially the same bonding pattern in the wires,
i.e. even-numbered wires are dimerized, and odd-numbered wires
additionally have a kink in the center. The distance A-1 between
apex atom and wire stays larger than the interatomic distances in
the wire. These distances can be modified by stretching or
compressing the wire, but the dimerization pattern is robust. In
conclusion, the optimized wire structures presented in
Table~\ref{tab2} can be considered as typical structures that are
formed by sodium finite chains suspended between two semi-infinite
electrodes.

\section{Conductance oscillation}\label{natres}
Our calculations of the conductance are based on the mode-matching
technique and we use a real-space finite-difference representation
of the Kohn-Sham Hamiltonian and the wave
functions.\cite{Khomyakov:prb04} As a first step, the one-electron
self-consistent potentials of the bulk leads and the scattering
region containing the wire are obtained from DFT calculations.
Subsequently the scattering problem is solved at the Fermi energy by
matching the modes in the leads to the wave function in the
scattering region. The conductance $G$ can be expressed in terms of
normalized transmission amplitudes $t_{n, n^{\prime}}$ using the
Landauer-B\"uttiker formula\cite{Buttiker:prb85}
\begin{equation}
G = G_0\, \sum_{n,n^{\prime}} | t_{n, n^{\prime}} |^{2},
\end{equation}
where $n$ and $n^{\prime}$ label the right-going modes in the left
and right leads, respectively and $G_0=e^2/\pi\hbar$. An efficient
implementation of a high-order finite-difference scheme for solving
the scattering problem is discussed in
Ref.~\onlinecite{Khomyakov:prb04}.
\begin{figure}[!tbp]
\includegraphics[width=7.0cm,keepaspectratio=true]{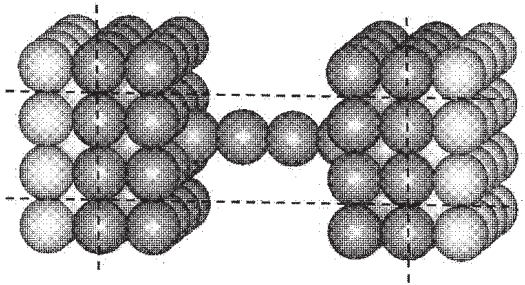}
\caption{(Color online) Structure of an atomic wire consisting of
two sodium atoms between two sodium leads terminated by (001)
surfaces. The boundaries of the supercell are indicated by dashed
lines. Bulk atoms are indicated by yellow (light grey) balls and
atoms in the scattering region by blue (dark grey) balls,
respectively.} \label{nawire}
\end{figure}

In more detail, the one-electron potentials of the leads and the
scattering regions are extracted from two DFT calculations for bulk
bcc sodium and for the supercell shown in Fig.~\ref{nawire},
respectively. For these calculations we use a local
Troullier-Martins pseudopotential\cite{Troullier:prb91} with a core
radius $r_{c} = 2.95\, a_{0}$; only the $3s$ electrons of sodium are
treated as valence electrons. All plane waves are included up to a
kinetic energy cutoff of $16$ Ry. We use $32^3$ $\mathbf{k}$-points
to sample the Brillouin zone (BZ) of the cubic \textit{bcc} unit
cell of bulk sodium. In our supercell calculations, $8^2$
$\mathbf{k}$-points are used to sample the lateral BZ in case of a
$2\times 2$ supercell, and $6^2$ $\mathbf{k}$-points in case of a
$3\times 3$ supercell. In all cases a Methfessel-Paxton smearing
with $\sigma=0.1$ eV is applied. Total energies are converged to
within $5\times 10^{-7}$ Hartrees.

One assumes that the leads outside the scattering region are
perfectly crystalline bulk material. So at the edges of the
scattering region, the potential should join smoothly to the
potentials of the bulk leads. We have checked that this is the case.
Enlarging the scattering region by including two extra atomic layers
in each lead changes the results reported for the conductance only
by $\sim 1.5\%$ for even-numbered wires and $\lesssim 0.5\%$ for
odd-numbered wires. The Fermi energy is extracted from the bulk
calculation.\cite{footnote:fermienergy} The only parameters in
calculating the conductance within the mode-matching
finite-difference scheme are the order $N$ of the finite-difference
approximation of the kinetic energy (i.e., the second derivative)
and the spacing $h_{x,y,z}$ between the real-space grid points. We
use $N=4$ and $h_{x,y,z} = 0.80a_0$; for details and convergence
tests we refer to Ref.~\onlinecite{Khomyakov:prb04}. The total
transmission is averaged over the $\mathbf{k}_{\|}$-point grid of
the lateral BZ of the supercell. To calculate the transmission it is
important to apply a proper $\mathbf{k}_{\|}$-point sampling. This
will be discussed in Appendix~\ref{kpoint}. Most calculations are
done for a $2\times 2$ lateral supercell. Enlarging the supercell
changes the conductance only marginally as will be discussed in
Sec.~\ref{contact}.

The electron transport in the crystalline leads is ballistic, i.e.
an electron goes through the leads without any scattering. The
transport properties of a monatomic wire suspended between two leads
depend upon three factors; the number of atoms in the chain, the
geometry of the wire, and the contact between wire and leads. In
Sec.~\ref{univbehav} we will discuss how these factors influence the
conductance. In the present section we will analyze the conductance
of monatomic sodium wires in a reference geometry, where all Na-Na
bond lengths are chosen to be equal to the bulk value $6.91a_0$, see
Table~\ref{tab1}. As in the previous section we attach a finite
atomic wire to the leads via two apex atoms, which then have a
coordination number $5$. All atoms in the wire have a coordination
number $2$.
\begin{figure}[!tbp]
\includegraphics[width=9.0cm,keepaspectratio=true]{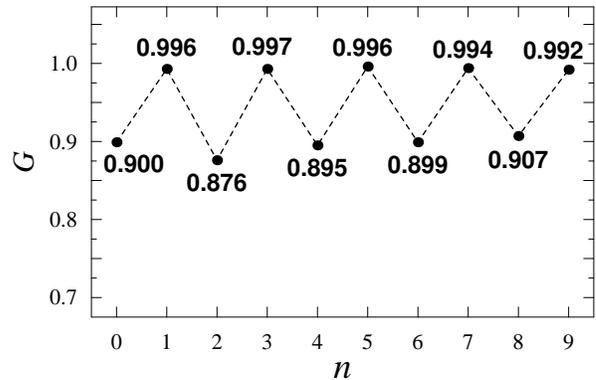}
\caption{Conductance (in units of $G_{0}$) as a function of the
number of atoms in the atomic chain. All atomic bond lengthes in the
system are equal to the bulk value $d=6.91a_{0}$.} \label{cond1}
\end{figure}

\subsection{First-principles calculations}
The calculated conductance as a function of the number of atoms in
the atomic chain is given in Fig.~\ref{cond1}. Since a sodium atom
has valence one, both the infinite sodium chain and bulk sodium have
a half-filled band, and the infinite wire has one conducting channel
at the Fermi level.\cite{footnote:fermilev_zgz} The conductance of
the infinite chain is then equal to the quantum unit $G_{0}$, and
the conductance of finite wires is $\leq G_{0}$. As can be observed
in Fig.~\ref{cond1} the conductance exhibits a regular oscillation
as a function of the number of atoms in the wire. The conductance is
very close to $G_{0}$ for odd-numbered wires, and for even-numbered
wires it is $\sim 10\%$ lower. Such a behavior of the conductance in
atomic-sized conductors is very different from ohmic behavior in
macroscopic conductors; it expresses the quantum nature of the
electron transport at the nanoscale.

\begin{figure}[!tbp]
\includegraphics[width=9.0cm,keepaspectratio=true]{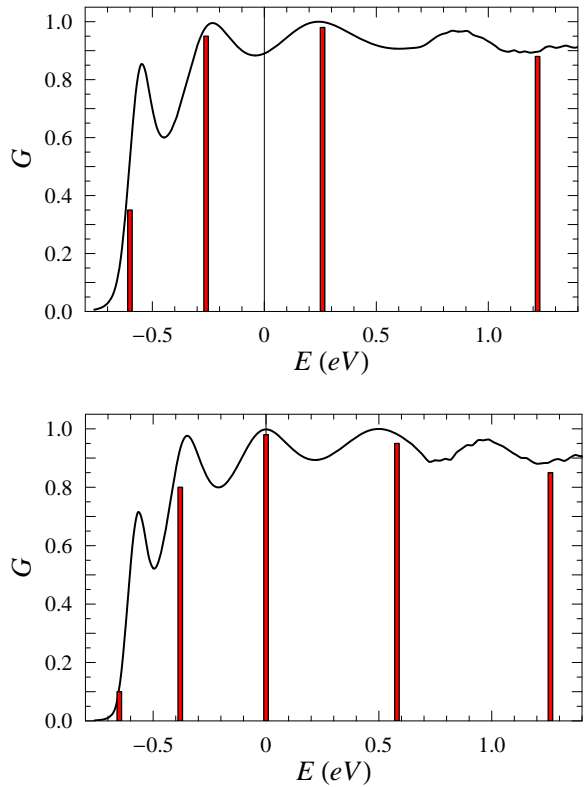}
\caption{(Color online) Conductance (in units of $G_{0}$) as a
function of energy for monatomic wires consisting of four (top
figure) and five (bottom figure) atoms. The red (grey) bars
correspond to the energy levels of free-standing wires. $E=0$
corresponds to the Fermi level.} \label{cond2}
\end{figure}

In order to interpret the even-odd oscillation we have calculated
the conductance as a function of energy for wires of different
length. The results for monatomic wires consisting of four and five
atoms are shown in Fig.~\ref{cond2}. Resonant peaks in the
conductance can be clearly identified. Qualitatively they correspond
to energy levels of a free-standing Na wire, which are shifted and
broadened into resonances by the interaction of the wire with the
leads. To illustrate this, the calculated energy levels of
free-standing wires of four and five atoms are shown as bars in
Fig.~\ref{cond2}. The levels are sufficiently close to the resonant
energies to warrant an interpretation of the conductance in terms of
a transmission through levels of the wire. As is clearly observed in
Fig.~\ref{cond2}, the Fermi level is in between two resonant peaks
for a four atom wire and right on top of a resonance for a five atom
wire. By calculating the conductance as a function of energy for
wires of different length it can be shown that this observation can
be generalized. The Fermi level is between resonances for
even-numbered wires and on top of a resonance for odd-numbered
wires.

\begin{figure}[!tbp]
\includegraphics[width=4.5cm,keepaspectratio=true]{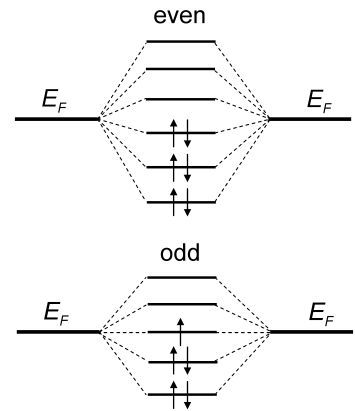}
\caption{Energy levels for odd- and even-numbered wires. The Fermi
level is in the middle of the HOMO-LUMO gap for even-numbered wires,
and it aligns with the HOMO level for odd-numbered wires.}
\label{tbm2}
\end{figure}

An intuitive picture of the transmission through the energy levels
is then presented by Fig.~\ref{tbm2}. Odd-numbered wires have a
highest occupied molecular orbital (HOMO) that is half filled.
Perfect transmission through this state takes place if the Fermi
level aligns with the HOMO. In even-numbered wires the HOMO is
completely filled and separated by a gap from the LUMO (lowest
unoccupied molecular orbital) level. The Fermi level is then in the
HOMO-LUMO gap. The position of the Fermi level with respect to the
HOMO level causes the off and on resonant behavior of the
conductance as a function of the wire length, which is causing a
regular even-odd oscillation of the conductance. In the next section
we will study this intuitive model in somewhat more detail by means
of a simple tight-binding model.

\subsection{Tight-binding model}\label{tbmodel}
\begin{figure}[!tbp]
\includegraphics[width=7.5cm,keepaspectratio=true]{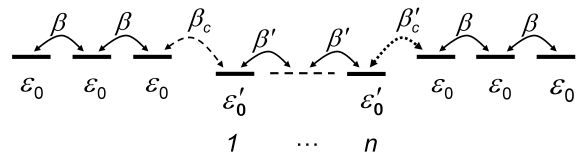}
\caption{Tight-binding representation of the $n$-atomic wire
attached to two semi-infinite one-dimensional leads.} \label{tbm}
\end{figure}

To support the intuitive picture presented in the previous section
we use a simple tight-binding model as shown in Fig.~\ref{tbm}, in
which the leads are modeled as quasi-one-dimensional systems
described by effective parameters. Here $\varepsilon_{0}$, $\beta$
are the on-site energies and nearest neighbor hopping coefficients
of the leads, and $\varepsilon_{0}^{\prime}$, $\beta^{\prime}$ are
the corresponding parameters of the wire. The coupling between the
left (right) electrode and the atomic chain is given by the hopping
coefficient $\beta_{c}$ ($\beta_{c}^{\prime}$).

If the system has mirror symmetry, the coupling is symmetric, i.e.
$\beta_{c} = \beta_{c}^{\prime}$. The leads and the chain are made
of the same material (sodium). If one assumes that all atoms are
neutral (local charge neutrality), then it is not unreasonable to
set $\varepsilon_{0} = \varepsilon_{0}^{\prime}$. The conductance
can be calculated analytically for this model by the mode-matching
technique.\cite{Khomyakov:prb05} The modes can be labeled by a wave
number $k$ in 1D Brillouin zone of the leads. The familiar relation
$E = \varepsilon_{0} + 2 \beta \cos(k a)$ gives for a half-filled
band the Fermi energy $E_F=\varepsilon_{0}$ and the Fermi wave
number $k_F=\pi/2a$. The parameter $\beta$ can be used as a scaling
parameter. In the following all energy parameters
$\varepsilon_{0},\varepsilon_{0}^{\prime},\beta^{\prime},\beta_{c},\beta_{c}^{\prime}$
are in units of $\beta$. The conductance of a wire at the Fermi
energy consisting of $n$ atoms is given by
\begin{eqnarray}\label{f01}
G &=& G_{0}, \quad \qquad \qquad \qquad \; n \rm{\, odd} \nonumber \\
  &=& G_{0}\;
 \frac{4\, \beta_{c}^4/\beta^{\prime 2}}{\left[ 1 +
\beta_{c}^4/\beta^{\prime 2} \right]^{2}}, \; \quad n \rm{\, even.}
\end{eqnarray}
The conductance for odd-numbered wires is equal to the quantum unit,
and it is smaller than the quantum unit for even-numbered wires
(unless $\beta_{c}^{2} = \beta^{\prime}$). This corresponds to the
situation shown in Fig.~\ref{tbm2}.

\begin{figure}[!tpb]
\begin{center}
$\begin{array}{l}
\mbox{\large\bf (a)} \\
\epsfxsize=8.0cm \epsffile{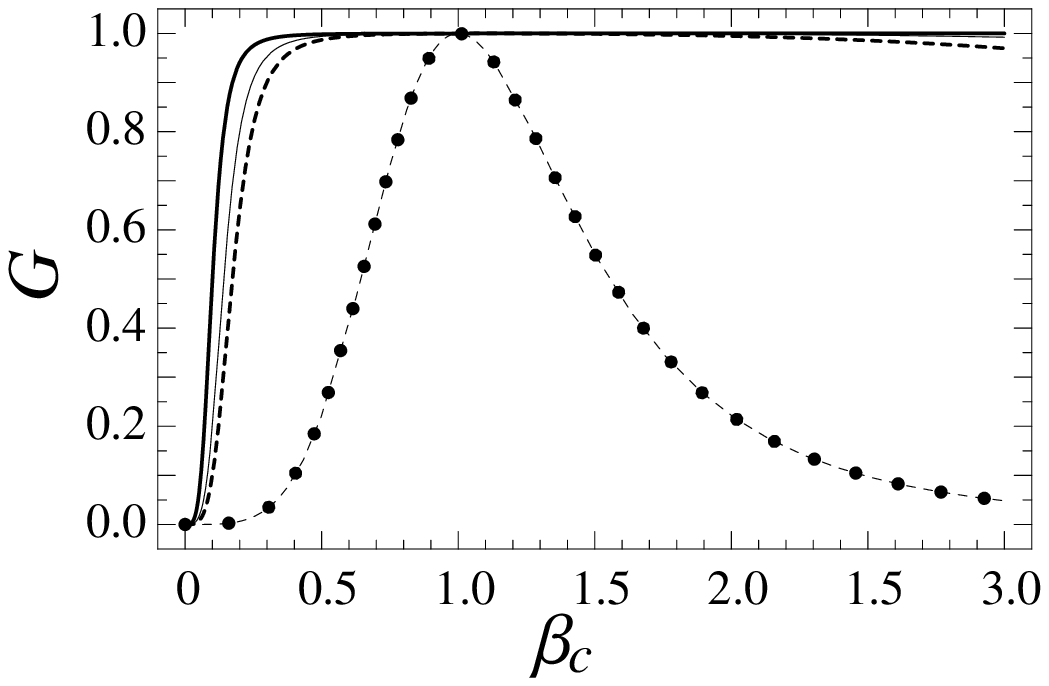} \\
\mbox{\large\bf (b)} \\
\epsfxsize=8.0cm \epsffile{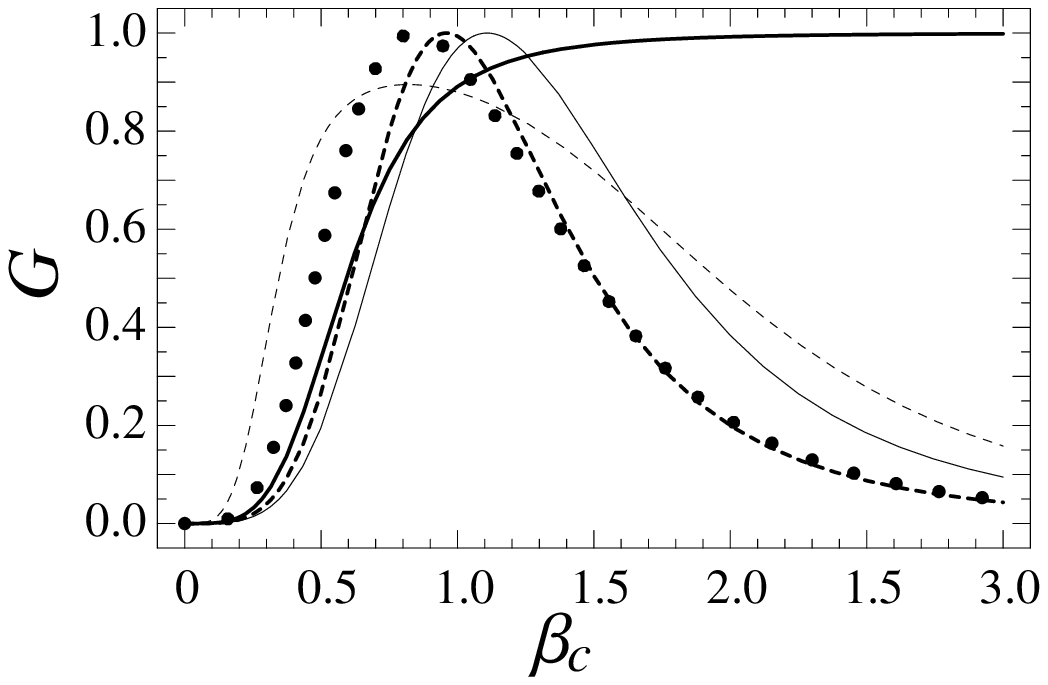}
\end{array}$
\end{center}
\caption{{\rm\bf (a)} Conductance (in units of $G_{0}$) at the Fermi
energy as a function of $\beta_{c}$ (in units of $\beta$) for wires
with $n=1$ (thick solid line), 2 (large dots), 3 (thin solid), 4
(thin dashed), and 5 (thick dashed) atoms; $\Delta\varepsilon =
0.02$; $\beta^{\prime} = 1$, {\rm\bf (b)} As {\rm (a)} with
$\Delta\varepsilon = 0.7$; $\beta^{\prime} = 1$.} \label{tbm3a}
\end{figure}

It is instructive to study some other consequences of the
tight-binding model. If $\Delta\varepsilon = \varepsilon_{0} -
\varepsilon_{0}^{\prime} \neq 0$ then a charge transfer will take
place between the leads and the wire. The conductance calculated at
the Fermi energy for a one-site wire ($n=1$) and a two-site wire
($n=2$) become, respectively,
\begin{eqnarray}
G &=& G_{0}\; \frac{4 \beta_{c}^{4}}{ \Delta\varepsilon^{2} + 4
\beta_{c}^{4}}, \label{f02}\\
G &=& G_{0}\; \frac{ 4 \beta_{c}^{4} \beta^{\prime 2}}{ \left[
\beta_{c}^{4} + \left( \beta^{\prime} + \Delta\varepsilon\right)^{2}
\right]
 \left[ \beta_{c}^{4} + \left(
\beta^{\prime} - \Delta\varepsilon\right)^{2} \right]  }.
 \label{f02b}
\end{eqnarray}
According to Eq.~(\ref{f02}) a nonzero $\Delta\varepsilon$
suppresses the transmission through a one-site wire. The
transmission is shifted ``off resonance'' and the conductance
becomes smaller than the quantum unit. However, the coupling between
wire and lead also causes a broadening of the resonance, which is
proportional to $\beta_{c}$. This broadening partially compensates
for the decrease of the conductance. If the coupling is sufficiently
strong, i.e. $4\beta_{c}^{4}\gg \Delta\varepsilon^{2}$, then the
conductance is again close to the quantum unit. In the limit of weak
coupling, i.e. $4\beta_{c}^{4}\ll \Delta\varepsilon^{2}$, the
conductance goes to zero with decreasing $\beta_{c}$ for any nonzero
$\Delta\varepsilon$. The conductance as a function of $\beta_{c}$ is
shown in Figs.~\ref{tbm3a}(a) and \ref{tbm3a}(b) for two different
values of $\Delta\varepsilon$.

The conductance of a two-site wire, see Eq.~(\ref{f02b}), behaves qualitatively different as a
function of the coupling strength $\beta_{c}$. In the weak coupling limit, i.e. $\beta_{c}^{4}\ll
(\beta^{\prime} \pm \Delta\varepsilon)^{2}$, corresponding to $\beta_{c}<1$ in Fig.~\ref{tbm3a}(a),
the conductance goes to zero with decreasing $\beta_{c}$ and the decrease is faster than for a
one-site wire. Note that this only holds for $\Delta\varepsilon \ll \beta^{\prime}$. If
$\Delta\varepsilon \sim \beta^{\prime}$ then the conductance decreases more slowly with decreasing
$\beta_{c}$ for a two-site wire than for a one-site wire, see the range $\beta_c<1$ in
Fig.~\ref{tbm3a}(b). If the coupling between wire and lead is strong, i.e. $\beta_{c}^{4}\gg
(\beta^{\prime} \pm \Delta\varepsilon)^{2}$, corresponding to $\beta_{c}>1$ in Figs.~\ref{tbm3a}(a)
and \ref{tbm3a}(b), then the conductance always decreases with increasing $\beta_{c}$. This is due
to a phenomenon called ``pair annihilation'' of resonances,\cite{Lee:prb01} which happens if the
resonance widths become larger than the spacing between the resonances. In a one-site wire this
cannot happen, since there is only one resonance. Between the strong and weak coupling regimes
there is a value of $\beta_{c}$ (close to 1) where the conductance of a two-site wire is equal to
the quantum unit, see Figs.~\ref{tbm3a}(a) and \ref{tbm3a}(b).

The conductance of longer wires, i.e. $n > 2$, can be interpreted
along the same lines. For small $\Delta\varepsilon$, the
odd-numbered wires resemble the one-site wire and the even-numbered
wires resemble the two-site wire, as shown in Fig.~\ref{tbm3a}(a).
For a very large range of coupling strengths $\beta_c$ one obtains
an even-odd oscillation in the conductance of a nearly constant
amplitude. The conductance of odd-numbered chains is close to the
quantum unit and that of even-numbered chains is smaller by an
amount that depends upon the coupling between wire and lead.
Apparently, this is the case that corresponds to the results of our
first-principles calculations, see Fig.~\ref{cond1}.

If $\Delta\varepsilon$ becomes larger, the conductance of all wires
as a function of $\beta_c$ becomes qualitatively similar to that of
the two-site wire, see Fig.~\ref{tbm3a}(b) (except the one-site
wire, of course). The amplitude and even the phase of the
conductance oscillation as a function of the wire length then
strongly depends upon the coupling $\beta_c$ of the wire to the
lead. For instance, if $\beta_c\lesssim 0.7$ in Fig.~\ref{tbm3a}(b),
the conductance of even-numbered wires is higher than that of
odd-numbered wires and all conductances are smaller than the quantum
unit. Note that if $\Delta\varepsilon$ is significant, it will be
accompanied by a significant charge transfer between wire and leads.
Whether this situation occurs can be studied by self-consistent
first-principles calculations.

\section{Stability of conductance oscillation}\label{univbehav}
First-principles calculations on a reference geometry give a regular
even-odd oscillation of the conductance as a function of the wire
length, as discussed in the previous section. The odd-numbered wires
have the highest conductance, close to the quantum unit $G_{0}$. The
simplified tight-binding model suggests that the geometry might
influence the amplitude and even the phase of the conductance
oscillation. In Sec.~\ref{geometry} we have shown that monatomic
sodium wires can have a linear, zigzag, or dimerized geometry,
depending upon the boundary conditions. In this section we study the
influence of the wire geometry upon the conductance. In particular,
we focus on the question of whether the phase of the even-odd
conductance oscillation is robust to modifications of the wire
geometry.

\subsection{Tension or compression of linear wires}\label{wirelength}
We consider a uniform tension or compression of the wire. As in the
previous section the Na-Na distance between atoms in the wire is
kept at a uniform value $d$ and the distance $d_{c}$ between the
apex atom and the wire is equal to $d$. As reference we use the
results shown in Fig.~\ref{cond1} where $d=6.91a_0$, which
corresponds to the bond length in bulk sodium. If we take the
equilibrium bond length $d=6.30a_0$ of the infinite linear wire as a
characteristic bond length, then $d=6.91a_0$ corresponds to a wire
under tension, i.e. a stretched wire. A wire with $d=5.82a_0$, which
corresponds to the equilibrium bond length of a Na$_2$ molecule, is
a wire under compression.

\begin{figure}[!tbp]
\includegraphics[width=8.5cm,keepaspectratio=true]{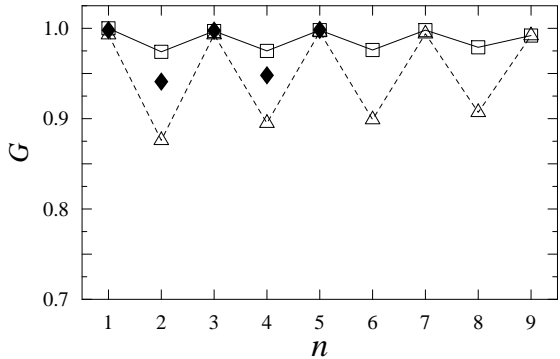}
\caption{Conductance (in units of $G_{0}$) as a function of the
number of atoms in the wire. The triangles correspond to stretched
wires with $d=6.91a_0$, squares to compressed wires with
$d=5.82a_0$, and diamonds to wires with $d=6.30a_0$.} \label{cond3}
\end{figure}

The results of first principles calculations of the conductance as a
function of the wire length for $d=5.82a_0$,$6.30a_0$,$6.91a_0$ are
presented in Fig.~\ref{cond3}. In all cases the conductance exhibits
a regular even-odd oscillation and the conductance of the
odd-numbered wires is close to the quantum unit. The conduction of
the even-numbered wires is smaller than the quantum unit and it
depends only weakly on the number of atoms in the wire. According to
the tight-binding analysis in Sec.~\ref{tbmodel}, this suggests that
the charge transfer between wire and lead, represented by
$\Delta\varepsilon$ in Eqs.~(\ref{f02}) and (\ref{f02b}), is very
small, see Fig.~\ref{tbm3a}(a).

The amplitude of the oscillation decreases with decreasing $d$. Two
opposing effects influence the conductance if we decrease the
interatomic spacing $d$ in the wire. Firstly, the spacing between
the resonant levels increases. In tight-binding terms the parameter
$\beta^{\prime}$ increases, which tends to decrease the conductance
of even-numbered wires, see Eq.~(\ref{f01}). Secondly, since
$d_c=d$, the resonances become broader if we decrease the distance
between the wire and the lead. Again in tight-binding terms the
parameter $\beta_c$ increases, which tends to increase the
conductance of even-numbered wires, cf.
Eq.~(\ref{f01}).\cite{footnote:pair_annih} According to
Fig.~\ref{cond3} the effect of the resonance broadening upon the
conductance is larger than the effect of increased resonance
spacing.

\begin{figure}[!tbp]
\includegraphics[width=8.0cm,keepaspectratio=true]{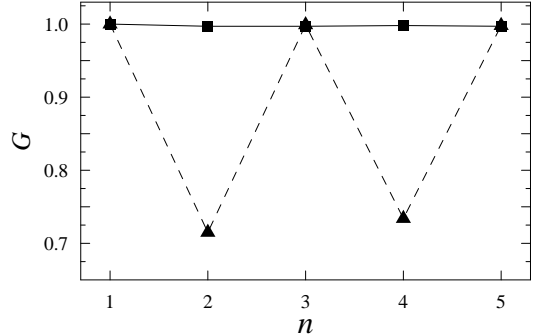}
\caption{Conductance (in units of $G_{0}$) as a function of the
number of atoms in the wire. The triangles correspond to wires with
$d=5.82a_0$ and $d_{c}=6.91a_0$ and the squares to wires with
$d=6.91a_0$ and $d_{c}=5.82a_0$.} \label{cond4}
\end{figure}

One can strengthen this analysis by varying the interatomic distance
$d$ in the wire, and the distance $d_c$ between the apex atom and
the wire independently. Figure~\ref{cond4} shows the calculated
conductance for a wire with $d<d_c$, i.e. $d=5.82a_0$ and
$d_{c}=6.91a_0$. The conductance oscillations are quite large, which
can be attributed to the increased resonance spacing discussed in
the previous paragraph. A small $d$ results in a large spacing
between the resonant levels of the wire. Therefore, the transmission
of even-numbered wires, which is off resonance at the Fermi level,
is low, whereas the transmission of odd-numbered wires stays on
resonance and is high.

If we calculate the conductance for a wire with $d>d_c$, i.e.
$d=6.91a_0$ and $d_{c}=5.82a_0$, we see in Fig.~\ref{cond4} that the
conductance oscillation is strongly suppressed. It can be attributed
to the resonance broadening. If the coupling between the wire and
the lead is strong, the resonances of the wire are wide. The
transmission in even-numbered wires is then relatively high, whereas
the transmission in odd-numbered wires stays close to the quantum
unit. Figure~\ref{cond4} shows that in the case of a strong coupling
between wire and lead the amplitude of the even-odd oscillation in
the conductance can become very small. According to the
tight-binding model, Eq.~(\ref{f01}), this happens if
$\beta_c^4/\beta^{\prime 2}\approx 1$. Note that such a strong
coupling is less likely for sodium monatomic wires with optimized
geometries, because the results discussed in Sec.~\ref{geometry}
indicate that $d<d_c$. According to Ref.~\onlinecite{Okamoto:prb99}
$d>d_c$ in gold monatomic chains, which might explain the small
amplitude of the conductance oscillation found experimentally in
gold wires.\cite{Smit:prl03}.

In conclusion, stretching or compressing the wire changes the
amplitude of the conductance oscillation, but it preserves its phase
and the value of the conductance for odd-numbered wires, which is
close to unity.

\subsection{Contact geometry}\label{contact}
\begin{figure}[!tbp]
\includegraphics[width=6.5cm,keepaspectratio=true]{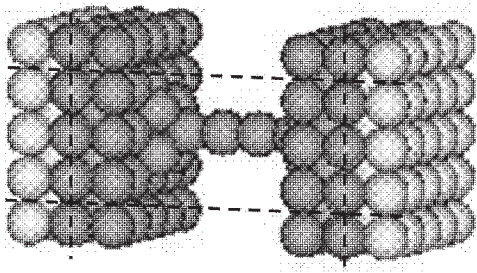}
\caption{(Color online) Structure of an atomic wire consisting of
two sodium atoms between two sodium leads terminated by (001)
surfaces. The atomic wire is connected to each surface via a five
atom pyramid. The boundaries of the $3\times 3$ supercell are
indicated by dashed lines. Bulk atoms are indicated by yellow (light
grey) balls and atoms in the scattering region by blue (dark grey)
balls, respectively.} \label{sharptip}
\end{figure}

\begin{figure}[!tbp]
\includegraphics[width=8.0cm,keepaspectratio=true]{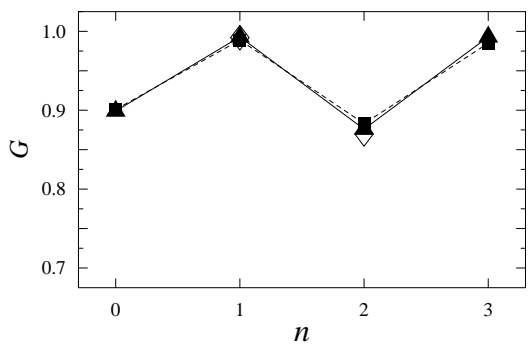}
\caption{Conductance (in units of $G_{0}$) as a function of the
number of atoms in the atomic wire. Squares refer to the $3\times 3$
supercell with five atom pyramid contacts, diamonds to the $3\times
3$ supercell with one apex atom contacts, and triangles to the
$2\times 2$ supercell with one apex atom contacts.} \label{cond5}
\end{figure}

\begin{figure}[!tpb]
\begin{center}
$\begin{array}{l}
\mbox{\large\bf (a)} \\
 \epsfxsize=7.5cm
\epsffile{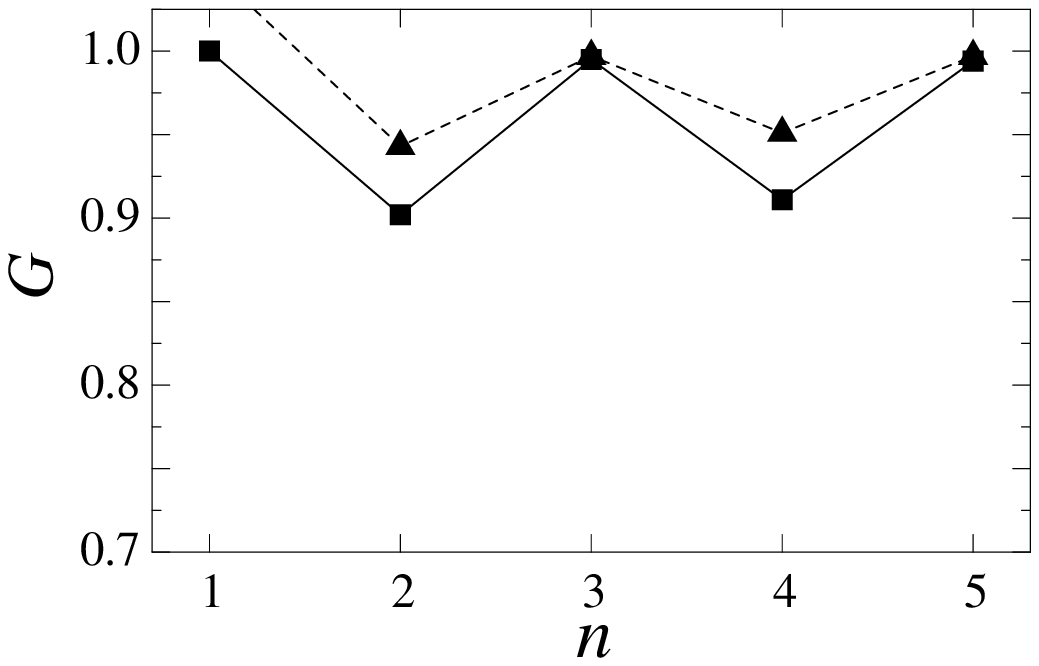} \\
\mbox{\large\bf (b)} \\
    \epsfxsize=7.5cm
    \epsffile{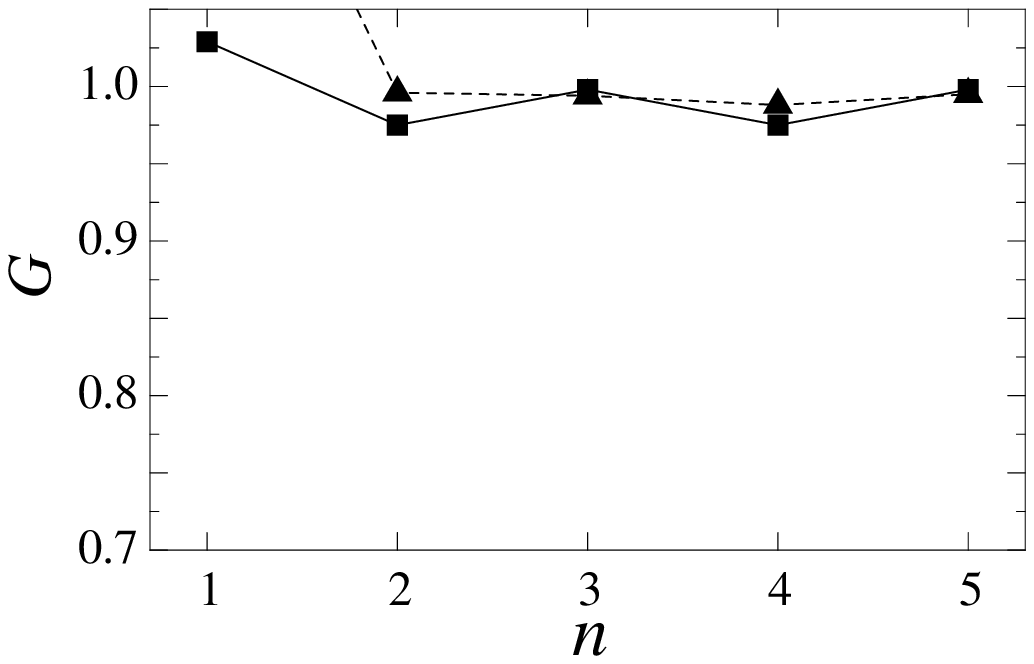}
\end{array}$
\end{center}
\caption{(a) Conductance (in units of $G_{0}$) as a function of the number of atoms in the atomic
wire with interatomic distance $d=d_c=6.91a_0$. The triangles (dashed line) correspond to
``direct'' contacts between wire and leads. The squares (solid line) correspond to a direct contact
at one end of the wire and a one apex atom contact at the other end. (b) As (a) with
$d=d_c=5.82a_0$.} \label{cond9}
\end{figure}

The coupling between the wire and the leads could be influenced by
the detailed geometry of the two contacts between the wire and the
leads.~\cite{Sim:prl01} Since the geometries of the wire-lead
contacts are not known from experiment, it makes sense to study the
sensitivity of the calculated conductances to these geometries. So
far in our calculations we have modeled both contacts by one apex
atom placed in a hollow site on the (001) surface in a $2\times 2$
lateral supercell. A more ``gradual'' contact is formed by a five
atom pyramid placed on (001) surface, as is shown in
Fig.~\ref{sharptip}. This requires using (at least) a $3\times 3$
supercell. To check that the size of the supercell does not
influence the results, we have also done calculations for a $3\times
3$ supercell with one apex atom contacts. The calculated
conductances are shown in Fig.~\ref{cond5}. The conductance of
monatomic sodium wires seems to be relatively insensitive to the
contact geometry. At the same time it shows that the results
obtained with the $2\times 2$ supercell are converged.

Another way of modifying the contacts is to remove the apex atoms
and position the first and the last atom of the wire on top of an
atom in the (001) surface layer. The calculated conductances are
shown in Figs.~\ref{cond9}(a) and \ref{cond9}(b) for the interatomic
distances $d=d_c=6.91a_0$ and $d=d_c=5.82a_0$, respectively. The
results obtained with these ``direct'' wire-surface contacts look
very similar to the ones obtained with one apex atom contacts, see
Fig.~\ref{cond3}. The amplitude of the even-odd oscillation is
somewhat smaller for the ``direct'' coupling. According to the
analysis presented in Sec.~\ref{wirelength} this indicates a
stronger coupling between wire and leads, or in tight-binding terms,
a larger $\beta_c$, cf. Eq.~(\ref{f01}). Note that the conductance
of one-atom wires in Figs.~\ref{cond9}(a) and \ref{cond9}(b) is
higher than $G_{0}$ due to direct tunneling between the electrodes.

One can also break the symmetry and use a direct contact between the
wire and one of the leads, and a one apex atom contact between the
wire and the other lead. The calculated conductances are given in
Figs.~\ref{cond9}(a) and \ref{cond9}(b). A comparison with symmetric
direct contacts and symmetric one apex atom contacts, see
Fig.~\ref{cond3}, shows that the phase of the even-odd conductance
oscillation is the same and the amplitude is in between that of the
two symmetric cases. It means that, besides the already mentioned
stronger coupling between wire and lead for the ``direct'' contact,
this symmetry breaking has little effect on the conductance.

We conclude that varying the geometries of the contacts between wire and leads does not have
a large effect on the regular even-odd oscillation of the conductance.

\subsection{Wire geometry: zigzag wires and dimerization}\label{dimstr}
\begin{figure}[!tbp]
\includegraphics[width=8.0cm,keepaspectratio=true]{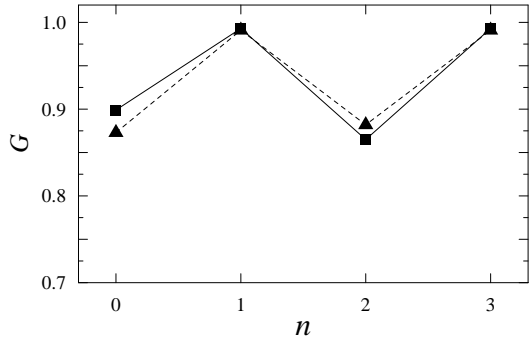}
\caption{Conductance (in units of $G_{0}$) as a function of the
number of atoms in the atomic wire. Dashed line and filled
rectangles correspond to zigzag wires; straight line and filled
squares to linear wires with $d=6.91a_{0}$.} \label{cond6}
\end{figure}

In Sec.~\ref{geometry} we studied the geometry of infinite zigzag
chains. In principle, a structural zigzag deformation could modify
the conductance of a finite monatomic
wire.\cite{Egami:ntech05,Sim:prl01}. Figure~\ref{cond6} shows the
calculated conductance of a wire with one apex atom contacts and
bond lengths $d=d_c=6.91a_0$ to which a zigzag distortion pattern is
applied with an amplitude corresponding to 15\% of the bond length.
 Compared to
straight wires, the conductance of zigzag even-numbered wires
changes by $\lesssim 3\%$, whereas the conductance of odd-numbered
wires is hardly affected at all. Such small effects are in line with
results reported previously.\cite{Egami:ntech05,Sim:prl01}

\begin{figure}[!tbp]
\includegraphics[width=8.0cm,keepaspectratio=true]{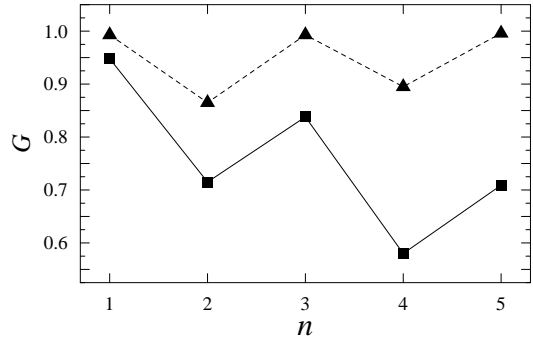}
\caption{Conductance (in units of $G_{0}$) as a function of the number of atoms in the atomic
wire. The squares correspond to dimerized wires with alternating bond lengths of $6.91a_0$
and $5.82a_0$; as a reference, the triangles correspond to wires with a uniform bond length
$6.91a_0$. } \label{cond8}
\end{figure}

According to the results obtained in Sec.~\ref{geometry} finite
straight wires with equidistant atoms can spontaneously break their
symmetry by dimerization. The conductance of optimized broken
symmetry structures is discussed in the next section. Here we study
the influence of an excessive symmetry breaking. We apply a regular
dimerization pattern to the wire, which consists of an alternation
between long and short bonds with bond lengths $d=6.91a_0$ and
$d=5.82a_0$, respectively. Continuing this pattern into the contacts
this means that even-numbered wires have short $d_c=5.82a_0$
contacts to both leads, whereas odd-numbered wires have one short
$d_c=5.82a_0$ contact and one long $d_c=6.91a_0$ contact. The
results are shown in Fig.~\ref{cond8}.

This curve can be analyzed using the tight-binding model introduced
in Sec.~\ref{tbmodel}. Assuming charge neutrality, i.e.
$\Delta\varepsilon = 0$, we need to generalize Eq.~(\ref{f01}) to
the case where the coupling $\beta_{c}$ to the left and
$\beta_{c}^{\prime}$ to the right leads are different. In addition,
dimerization in the wire leads to an alternation of two hopping
coefficients $\beta^{\prime}$ and $\beta^{\prime\prime}$. The
conductance of an $n$-atomic wire is then given by
\begin{eqnarray}
G &=& G_0\; \frac{4\, \left(\frac{\beta_c}{\beta_c^\prime}\right)^2
\left(\frac{\beta^{\prime\prime}}{\beta^\prime}\right)^{n-1}
}{\left[ 1 + \left(\frac{\beta_c}{\beta_c^\prime}\right)^2
\left(\frac{\beta^{\prime\prime}}{\beta^\prime}\right)^{n-1}
\right]^2}, \;\; \quad n \, \mathrm{odd} \nonumber \\
  &=& G_{0}\; \frac{ 4\,
\left(\frac{\beta_{c}\beta_{c}^{\prime}}{\beta^{\prime}}\right)^{2}
\left(\frac{\beta^{\prime\prime}}{\beta^{\prime}}\right)^{n-2}
}{\left[ 1 +
\left(\frac{\beta_{c}\beta_{c}^{\prime}}{\beta^{\prime}}\right)^{2}
\left(\frac{\beta^{\prime\prime}}{\beta^{\prime}}\right)^{n-2}
\right]^{2}}, \quad n\, \mathrm{even}. \label{f04}
\end{eqnarray}

One notices from Eq.~(\ref{f04}) that even the shortest odd-numbered
wire, $n=1$ has a conductance smaller than the quantum unit if
$\beta_{c} \neq \beta_{c}^{\prime}$. This is observed in our first
principles results, where the conductance of the $n=1$ wire is
$0.95G_0$, see Fig.~\ref{cond8}. Furthermore, Eq.~(\ref{f04}) shows
that the conductance of both even-numbered and odd-numbered wires
decreases as a function of increasing $n$ if $\beta^{\prime} \neq
\beta^{\prime\prime}$. Also this is clearly observed in our first
principles calculations. Dimerization of an infinite wire creates a
gap in the density of states, so, in general, one expects that the
conductance drops as a function of the wire length. A decreasing
conductance for longer wires has been observed experientially for
platinum, but its nature has not been clarified
yet.\cite{Smit:prl03}

Finally, although the conductance for even- and odd-numbered wires
decreases as a function of wire length, its even-odd oscillation is
preserved. Assuming $\beta^{\prime\prime}/\beta^{\prime}=
\beta_{c}/\beta_{c}^{\prime}=x$ for odd-numbered wires and
$\beta^{\prime\prime}/\beta^{\prime}= \beta_{c}/\beta =
\beta_{c}^{\prime}/\beta^{\prime}=y$ for even-numbered wires the
tight-binding model, Eq.~(\ref{f04}), can be fitted to the
first-principles results. This yields the parameter ratio's
$x\approx 0.81$ and $y\approx 0.76$ for the odd- and even-numbered
dimerized wires presented in Fig.~\ref{cond8}.

\subsection{Optimized geometry}\label{optstr}
\begin{figure}[!tbp]
\includegraphics[width=8.0cm,keepaspectratio=true]{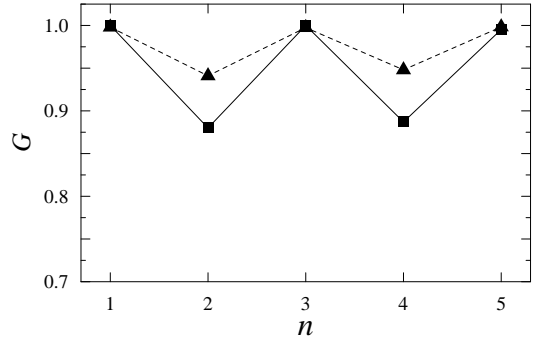}
\caption{Conductance (in units of $G_{0}$) as a function of the
number of atoms in the atomic wire. The triangles correspond to
linear wires with $d=6.30a_0$ and the squares to wires with
optimized geometry. } \label{cond7}
\end{figure}

In previous sections we studied the influence of the structure of a
monatomic wire upon its conductance by varying interatomic distances
corresponding to values ranging from the Na dimer $5.82a_0$ to the
Na bulk $6.91a_0$ values. In this section we discuss the conductance
for wires with optimized geometries, which were obtained in
Sec.~\ref{geometry}. Figure~\ref{cond7} shows the results from the
first-principles calculations. As a reference, it also shows the
results for wires with equidistant atoms corresponding to the
geometry of an infinite wire.

It can be observed that the amplitude of the conductance oscillation
for wires with optimized geometry is larger than for the reference
wires. This results from a slight decoupling of the wire from
electrodes, since for the optimized wires $d_{c} > \bar{d}$, see
Table~\ref{tab2} and the discussion in Sec.~\ref{contact}. As can be
seen from Fig.~\ref{cond7}, the phase of the even-odd conductance
oscillation is also observed for optimized geometries. Moreover,
despite the presence of a topological defect in the center of the
odd-numbered wires, the conductance of odd-numbered wires is close
to the quantum unit. This is actually predicted by the tight-binding
model for any odd-numbered wire whose geometry has mirror symmetry
with respect to a plane through the center and perpendicular to the
wire, provided $\Delta\varepsilon=0$.

One might expect the conductance for even-numbered wires to decrease
with the length of the wire, due to the effect of dimerization as
discussed in the previous section. This does not show in
Fig.~\ref{cond7}, because the dimerization in the optimized geometry
is much weaker. Therefore, the effect will show up only in wires
that are much longer.

\section{Beyond linear response}\label{currvolt}
In the linear response regime the current and therefore the
conductance are fully determined by the electrons at the Fermi
energy. If a finite bias $V$ is applied then the current is given by
\begin{equation}\label{f06}
I = G_{0}\, \int_{E_{\rm F} - V/2}^{E_{\rm F} + V/2}  T\left( E, V \right) \, d E ,
\end{equation}
where the transmission coefficient $T\left( E, V \right)$ depends on
the energy of the electron $E$ and the voltage $V$. The differential
conductance is defined as $G(V)=dI/dV$.

\begin{figure}[!tbp]
\includegraphics[width=8.0cm,keepaspectratio=true]{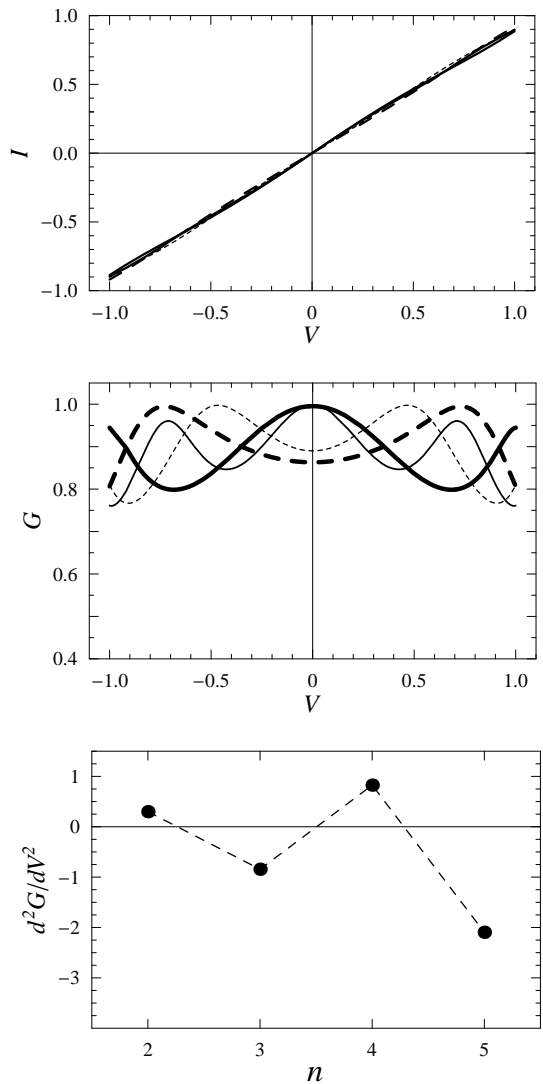}
\caption{Current-voltage characteristics for two (thick dashed
line), three (thick line), four (thin dashed line) and five (thin
line) atomic wires with $d=d_{c}=6.91a_{0}$. The top, middle and
bottom figures contain $I$-$V$ curves, the differential conductance
$G(V)=dI/dV$ as a function of $V$ and the second derivative of the
conductance $d^{2}G(0)/d^{2}V$ as a function of the number of atoms,
respectively.} \label{ivcurves}
\end{figure}

In this section we discuss some of the consequences of a finite
bias. We make the approximation $T\left( E, V \right)\approx T\left(
E \right)$, which is valid for a relatively small voltage in the
limit that the electronic structure of the wire is not changed by
the voltage. Examples of transmissions as a function of energy are
given in Fig.~\ref{cond2} for atomic wires with the ``reference''
geometry $d=d_{c}=6.91a_0$. The corresponding \textit{I-V} curves,
calculated from Eq.~(\ref{f06}), are given in Fig.~\ref{ivcurves}.
From the \textit{I-V} curves we calculate the differential
conductance and the second derivative of the conductance, which are
also presented in Fig.~\ref{ivcurves}.

The conductance varies by less than 5\% for biases up to $\sim \pm
0.2$V, which one might call the linear response regime. The
conductance decreases monotonically for odd-numbered wires and it
increases monotonically for even-numbered wires for biases up to
$\sim 0.5$V. The oscillating behavior of the conductance at higher
biases results from the resonant peaks in the transmission. At
biases larger than $\sim 0.5$V the non-self-consistent procedure
probably becomes increasingly inaccurate.\cite{Tsukamoto:prb02}

An important characteristic of the conductance curve that can be
measured experimentally, is its second derivative at the Fermi
energy. If the transmission near the Fermi energy $E_{\rm F}=0$ is
approximated by a polynomial function $T(E)=T(0) +
T(0)^{\prime\prime}\, E^{2}$, then the differential conductance is
$G(V) = G_{0}\,[T(0) + T(0)^{\prime\prime}\, V^{2}/4]$. In
Sec.~\ref{natres} we interpreted the even-odd conductance
oscillation of the conductance in terms of switching between off-
and on-resonance behavior. If this is true then the second
derivative of the conductance $d^{2}G/d^{2}V =
T(0)^{\prime\prime}/2$ must be positive for even-numbered wires and
negative for odd-numbered wires at the Fermi energy, as shown in
Fig.~\ref{ivcurves}. At the same time the first derivative is zero.

\section{Discussion}\label{otherworks}
In this section we compare our results to those obtained in previous
studies. The oscillating behavior of the conductance of monatomic
sodium wires was first suggested from calculations using planar
jellium electrodes.\cite{Lang:prl97} The conductance of both even-
 and odd-numbered wires is then significantly lower than the quantum
unit and the conductance of even-numbered wires is larger than that
of odd-numbered ones. Both of these features are likely to be
artifacts of using jellium electrodes, since adding atomic bases
between wire and jellium leads reverses the phase of the conductance
oscillation and makes the conductance of odd-numbered wires approach
the quantum unit.\cite{Kobayashi:prb00,Tsukamoto:prb02} Calculations
using tip-shaped jellium electrodes predict that the phase of the
conductance oscillation critically depends upon the sharpness of the
tips,\cite{Havu:prb02} although this effect is disputed in recent
calculations.\cite{Havu:prb06}

Using atomistic electrodes with a bulk structure our calculations
show that the conductance has a regular even-odd oscillation, in
which the conductance of the odd-numbered wires in optimized
structures is close to the quantum unit and that of even-numbered
wires is approximately 10\% lower. Only the latter is modified
substantially if the geometry of the wire or the contacts between
wire and electrodes are changed within reasonable bounds. Apparently
jellium electrodes cause reflections of electrons trying to enter
the wire, which results in the artifacts discussed above. The
amplitude of the conductance oscillation we find, is smaller by
$\sim 67\%$ than that obtained using jellium electrodes plus atomic
bases. This would indicate that atomic bases do not completely
remove the reflections caused by the jellium electrodes.

Calculations based upon one-dimensional metal electrodes predict
that it is possible to change the phase of the conductance
oscillation by varying the coupling between the wire and the
electrodes.\cite{Gutierrez:appb01} We did not observe such an effect
for three-dimensional atomistic electrodes. In a recent calculation
it is found that the conductance of odd-numbered wires decreases
sharply with increasing wire length, and even the phase of the
conductance oscillation can be reversed in long
wires.\cite{Major:cond05} This is not confirmed by our calculations,
where the conductance of odd-numbered wires stays close to the
quantum unit and the phase of the oscillation is stable. Experiments
on the even-odd conductance oscillation in monovalent gold wires do
not reveal a decrease of the conductance in odd-numbered wires, and
the conductance stays close to the quantum unit.\cite{Smit:prl03}

Other studies seem to indicate that the phase of the even-odd
conductance oscillation does not depend very sensitively upon the
structure of the electrodes, since the same phase is observed in
calculations using \textit{bcc} electrodes oriented in the (111)
direction,\cite{Sim:prl01} and in calculations using electrodes with
an artificial \textit{fcc} structure.\cite{LeeYJ:prb04} The
amplitude of the oscillation is much more sensitive, however. A
previous study on sodium wires suspended between sodium electrodes
gives an amplitude of only 1\%, which is an order of magnitude
smaller than what we find using similar
geometries.\cite{Egami:prb05} Although we do not know what the cause
of this difference is, we observe that the amplitude of the even-odd
conductance oscillation is sensitive to the one-electron potential
used in solving the scattering problem. This potential is obtained
from a self-consistent electronic structure calculation. Such
calculations frequently use a convergence criterion applied to the
total energy. However, since a variational principle does not apply
to the one-electron potential, the convergence criterion should be
much stricter in order to converge the potential.

\begin{figure}[!tbp]
\includegraphics[width=8.0cm,keepaspectratio=true]{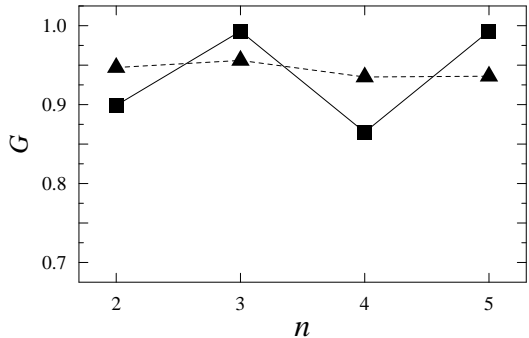}
\caption{Conductance (in units of $G_{0}$) as a function of the
number of atoms in the atomic wire. Triangles and squares correspond
to using a potential obtained from a total energy calculation with
energy convergence criterion set to $\Delta E=2\times 10^{-5}$ and $
5\times 10^{-7}$ Hartrees, respectively.} \label{condconv}
\end{figure}

As an illustration, Fig.~\ref{condconv} shows the conductance of sodium wires calculated from a
potential obtained with the usual energy convergence criterion, compared to one obtained with a
stricter energy convergence criterion. We have checked that the result does not change anymore if
the convergence criterion is even made stricter. This figure clearly shows that changes in the
potential that are caused by small charge transfers can markedly influence the amplitude of the
even-odd conductance oscillation.

\section{Summary and Conclusions}
We have performed first-principles calculations to study the
stability of even-odd conductance oscillations in a sodium monatomic
wire with respect to structural variations. An infinite sodium wire
can be linear and consists of equidistant atoms or dimers of atoms,
or it can have a zigzag structure, depending upon the tension or
compression applied to the wire. The geometry of finite sodium
wires, suspended between two sodium electrodes, is influenced by
boundary effects. Wires comprising an even or odd number of atoms
are dimerized, but odd-numbered wires have a topological defect in
the center.

In the linear response regime the conductance is determined by the
electrons at the Fermi energy. The conductance of sodium wires shows
a distinct even-odd oscillation. The odd-numbered wires have a
conductance close to the quantum unit $G_0=e^2/\pi\hbar$ and
even-numbered wires have a lower conductance. This oscillation is
remarkably robust, as we show by systematically varying the
structure of the wires and the geometry of the contacts between the
wires and the electrodes. The phase of the oscillation is not
affected by these structural variations, i.e. odd-numbered wires
have a higher conductance than even-numbered ones. Moreover,
odd-numbered wires have a conductance close to the quantum unit,
unless the structural deformation of the wire becomes very large and
the contact to the left lead is markedly different from that to the
right lead. The conductance of even-numbered wires is much more
sensitive to the wire geometry. Increasing the interatomic distances
in the wire and/or strengthening the contacts between wire and leads
increases the conductance of even-numbered wires; increasing the
asymmetry between the interatomic distances or between left and
right contacts decreases the conductance.

These results can be interpreted on the basis of resonant
transmission. For odd-numbered wires the Fermi energy coincides with
a resonance in the transmission, whereas for even-numbered wires the
Fermi energy is between two resonances. Changing the geometry of the
wire or the contacts affects the spacing between the resonances and
their widths and therefore it affects the conductance of
even-numbered wires; decreasing the spacing and/or increasing the
widths increases the conductance.\cite{footnote:pair_annih} Since
for odd-numbered wires the Fermi level is pinned at a resonance,
their conductance is affected much less by changing the wire
geometry.

We have formulated a simple tight-binding model to analyze these
results. It shows that the even-odd conductance oscillation is
stable with respect to structural variations, unless the on-site
energies for atoms in the wire are substantially different from the
on-site energies of atoms in the leads. Note that a large difference
in on-site energies is necessarily accompanied by a significant
charge transfer between the wire and the leads. The results of the
first principles calculations demonstrate that this is not the case.
For wires with equidistant atoms that have mirror symmetry with
respect to a plane perpendicular to the wire, Eq.~(\ref{f01}) shows
that if all on-site energies are identical, the conductance of
odd-numbered wires is one quantum unit, whereas that of
even-numbered wires is determined by the ratio of the wire-lead
coupling and the atom-atom coupling within the wire. Breaking the
mirror symmetry, Eq.~(\ref{f04}) shows that the conductance of
odd-numbered wires becomes smaller than one unit. The symmetry
breaking has to be large, however, in order to have a sizable effect
on the conductance.

We have also calculated the current-voltage characteristics of
sodium wires in the low bias regime. The differential conductance
clearly shows a nonmonotonic behavior. In particular, the second
derivative of the conductance has an alternating sign as a function
of the number of atoms in the wire; even-numbered wires have a
positive second derivative and odd-numbered wires a negative one.
This effect can be ascribed to the resonant nature of the
transmission. It could be used to establish the resonant behavior of
the even-odd conductance oscillation experimentally.

Comparison to other work shows that simple jellium electrodes do not
reproduce the even-odd conductance oscillation correctly. Using
atomic bases yields the correct phase of the oscillation. The
conductance of odd-numbered wires is rather stable with respect to
varying the atomic structure, but that of even-numbered wires is
sensitive to structural details and the quality of the one-electron
potential.

\acknowledgments

This work was financially supported by the ``Nederlandse Organisatie
voor Wetenschappelijk Onderzoek (NWO)'' via the research programs of
``Chemische Wetenschappen (CW)'' and the ``Stichting voor
Fundamenteel Onderzoek der Materie (FOM)'', and by ``NanoNed'', a
nanotechnology programme of the Dutch Ministry of Economic Affairs.
Part of the calculations were performed with a grant of computer
time from the ``Stichting Nationale Computerfaciliteiten (NCF)''.

\appendix
\section{$\mathbf{k}$-point sampling}\label{kpoint}

\begin{figure}[!tbp]
\includegraphics[width=9.3cm,keepaspectratio=true]{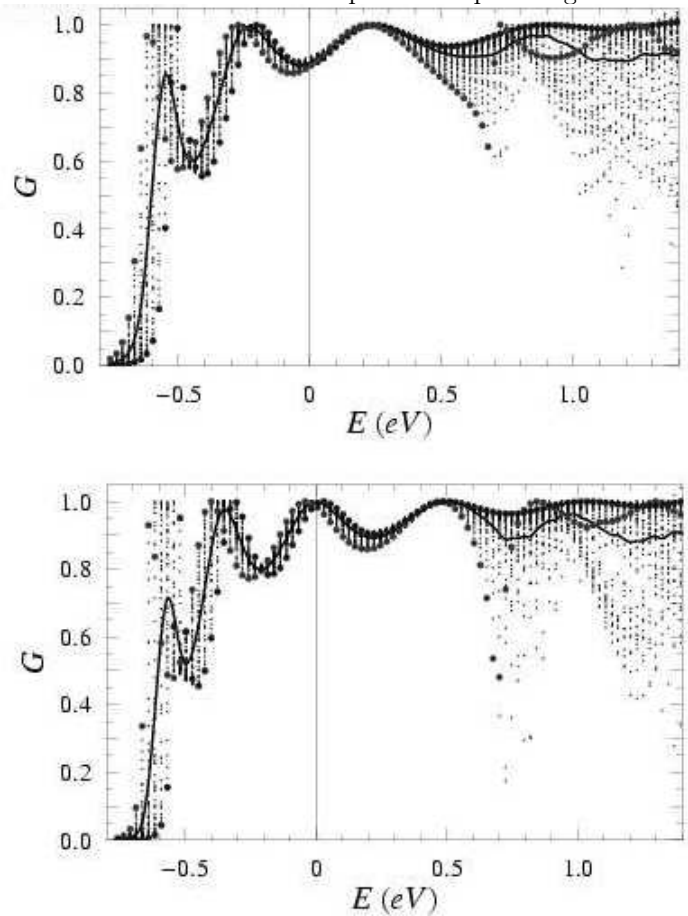}
\caption{(Color online) Conductance (in units of $G_{0}$) as a
function of energy for four (top) and five (bottom) atomic wires.
The dots correspond to the conductance for various
$\mathbf{k}_{\parallel}$-points of a $20\times 20$ grid in the BZ
($55\, \mathbf{k}_{\parallel}$-points in the irreducible BZ). The
thick line is the averaged conductance. The red (light grey) and
blue (dark grey) dots correspond to $\mathbf{k}_{\parallel} = (0,0)$
and $(0.5,0.5)$, respectively.} \label{kpoints}
\end{figure}

In this Appendix we discuss the effect of
$\mathbf{k}_{\parallel}$-point sampling on the conductance. In
modeling a conductor between two semi-infinite electrodes one
usually assumes a supercell geometry in the lateral direction. The
scattering region then consists of a periodic array of parallel
wires, and the lateral supercell must be chosen large enough to
prevent an interaction between these wires. To limit the
computational demands the supercell is chosen as small as possible,
without a significant loss of accuracy. According to the results
obtained in Sec.~\ref{contact}, using a $2\times 2$ supercell is
already sufficient for the Na system discussed here. We average the
conductance over the 2D Brillouin zone (BZ)
\begin{equation}
G = \frac{1}{N_{\parallel}} \sum_{\mathbf{k}_{\parallel}}\, G_{\mathbf{k}_{\parallel}},
\end{equation}
where $N_{\parallel}$ is the number of $\mathbf{k}_{\parallel}$-points used for the BZ
sampling. Calculating the conductance for an infinitely large supercell would include
contributions from off-diagonal transmission amplitudes between different
$\mathbf{k}_{\parallel}$. From our results we conclude that their contribution is small as
compared to the contribution of the diagonal terms $G_{\mathbf{k}_{\parallel}}$ already for a
$2\times 2$ supercell.

The calculated conductance $G_{\mathbf{k}_{\parallel}}$ as a
function of ${\mathbf{k}_{\parallel}}$ for four and five atomic
wires is shown in Fig.~\ref{kpoints}. The dispersion of the
conductance is relatively small around the Fermi energy, which means
that using a coarse ${\mathbf{k}_{\parallel}}$ grid to calculate the
conductance in the linear response regime is reasonable. The results
discussed in sections have been obtained using a $6\times 6$ grid in
the BZ ($6\, \mathbf{k}_{\parallel}$-points in the irreducible BZ).
Fig.~\ref{kpoints} shows that even sampling the BZ with a single
$\mathbf{k}_{\parallel}$-point can give a reasonable result. This
can be accidental, however, since the figure also demonstrates that
the dispersion is quite large both for energies lower and for
energies higher than the Fermi energy. Especially for higher
energies a single ${\mathbf{k}_{\parallel}}$-point is clearly
insufficient for calculating the conductance. This regime becomes
important if current-voltage characteristics are calculated, because
such calculations require an integration over a wide energy range.
The large dispersion at higher energies is related to an increased
number of van Hove singularities in the leads. This suggests that
$\mathbf{k}_{\parallel}$-point sampling can be important for leads
containing atoms with a valency higher than 1, because the number of
van Hove singularities is then usually also higher due to a more
complicated band structure.\cite{Thygesen:prb05a}


\begin{thebibliography}{75}
\bibitem{footnote:fermienergy} We follow the procedure outlined in Ref.~\onlinecite{Khomyakov:prb04}.
to define the Fermi energy for the conductance calculations. In
connecting the leads to the scattering region the average potential
of the bulk leads is aligned to the average potential in the middle
of the supercell slab. The Fermi energies of bulk sodium and of the
supercell are measured relative to their respective average
potentials. If the slab is sufficiently thick, this procedure
ensures an alignment of the Fermi energies of the bulk leads and of
the scattering region. In our calculations we have found a numerical
difference between the bulk and supercell Fermi energies in the
range $1\div 15$ meV. We have verified explicitly that the
conductance is not sensitive to a shift of $\sim 20$ meV of the
Fermi energy, see e.g. Fig.~\ref{kpoints}.
\bibitem[{foo({\natexlab{a}})}]{footnote:fermilev_zgz}
\bibinfo{note}{The Fermi level can cross two energy bands for a strongly
  deformed chain with a zigzag structure. In this case the chain has two
  conducting channels.}
\bibitem[{foo({\natexlab{b}})}]{footnote:pair_annih}
\bibinfo{note}{This conclusion is based on an additional assumption that
  resonances around the Fermi level are not annihilated by a strong coupling
  between the wire and the leads. Otherwire, the conductance for even-numbered
  wires decreases if we increase the coupling parameter $\beta_{c}$ by decreasing the
  distance between the wire and the lead.}
\end{thebibliography}

\end{document}